\documentclass[journal]{IEEEtran}
\usepackage[utf8x]{inputenc}
  \usepackage[dvipdf]{graphicx}
  \graphicspath{{figures/}}
  \usepackage{multirow}
  \usepackage[caption=false]{subfig}
  \DeclareGraphicsExtensions{.pdf,.jpeg,.png}
\usepackage{xcolor}

\usepackage{amsmath}
\usepackage[linesnumbered,ruled,vlined]{algorithm2e}

\SetCommentSty{mycommfont}

\usepackage{stfloats}
\usepackage{comment} 
\usepackage{tabularx,booktabs}
\newcolumntype{Y}{>{\centering\arraybackslash}X}
\usepackage{arydshln}
\newcommand{\midsepremove}{\aboverulesep = 0mm \belowrulesep = 0mm} \newcommand{\midsepdefault}{\aboverulesep = 0.605mm \belowrulesep = 0.984mm}

\makeatletter
\newcommand{\removelatexerror}{\let\@latex@error\@gobble}
\makeatother

\begin{document}

% \title{Classification, Regression and Clustering ML Methods For Real-time Network Fault Monitoring of Quasi-Periodic IoT Devices \& Networks}
\title{Machine Learning Methods for Monitoring of Quasi-Periodic Traffic in Massive IoT Networks}
% \author[1]{René~B.~Sørensen,~\IEEEmembership{Student Member,~IEEE}}
% \author[2]{Jimmy~J.~Nielsen,~\IEEEmembership{Member,~IEEE}}
% \author[3]{Petar~Popovski,~\IEEEmembership{Fellow,~IEEE}}
% \affil[1,2,3]{Aalborg University, Aalborg, Denmark}
% \affil[ ]{\{petarp, jjn, rbs\}@es.aau.dks}
\author{René~B.~Sørensen,~\IEEEmembership{Student Member,~IEEE}
Jimmy~J.~Nielsen,~\IEEEmembership{Member,~IEEE} 
and Petar~Popovski,~\IEEEmembership{Fellow,~IEEE} 
\thanks{René B. Sørensen, Jimmy J. Nielsen and Petar Popovski are with the Connectivity section within the department
of Electronic Systems, Aalborg University, Denmark,
 e-mail: (rbs, jjn, petarp)@es.aau.dk.}% <-this % stops a space
\thanks{Manuscript submitted December 20, 2019.}}
%         John~Doe,~\IEEEmembership{Fellow,~OSA,}
%         and~Jane~Doe,~\IEEEmembership{Life~Fellow,~IEEE}% <-this % stops a space
% \thanks{M. Shell was with the Department
% of Electrical and Computer Engineering, Georgia Institute of Technology, Atlanta,
% GA, 30332 USA e-mail: (see http://www.michaelshell.org/contact.html).}% <-this % stops a space
% \thanks{J. Doe and J. Doe are with Anonymous University.}% <-this % stops a space
% \thanks{Manuscript received April 19, 2005; revised August 26, 2015.}


\markboth{IEEE Internet of Things Journal, 2020}%
{René B. Sørensen \MakeLowercase{\textit{et al.}}: A Passive, Parametric Machine Learning Approach to Real-time Network Fault Monitoring for Quasi-Periodic Devices \& Networks}



\maketitle

\textit{©2020 IEEE.  Personal use of this material is permitted. Permission from IEEE must be obtained for all other uses, in any current or future media, including reprinting/republishing this material for advertising or promotional purposes, creating new collective works, for resale or redistribution to servers or lists, or reuse of any copyrighted component of this work in other works.}

\begin{abstract}
%The Internet of Things (IoT) has matured to the point that multiple wireless interfaces are available for IoT connectivity. 
One of the central problems in massive Internet of Things (IoT) deployments is the monitoring of the status of a massive number of links. The problem is aggravated by the irregularity of the traffic transmitted over the link, as the traffic intermittency can be disguised as a link failure and vice versa. 
In this work we present a traffic model for IoT devices running quasi-periodic applications and we present unsupervised, parametric machine learning methods for online monitoring of the network performance of individual devices in IoT deployments with quasi-periodic reporting, such as smart-metering, environmental monitoring and agricultural monitoring. Two clustering methods are based on the Lomb-Scargle periodogram, an approach developed by astronomers for estimating the spectral density of unevenly sampled time series. 
%All methods require a minimal set of network meta-data, which can be sampled passively, and are tolerant of missing data (outage) and jitter (random delay). 
%Prior information about the monitored system changes the nature of the fault-monitoring problem. Each proposed method solves one of the three sub-problems: %Depending on the availability of prior information, The proposed methods each solve one of three sub-problems in fault-detection:
%labeling, regression, and clustering.
We present probabilistic performance results for each of the proposed methods based on simulated data and compare the performance to a na\"{i}ve network monitoring approach. The results show that the proposed methods are more reliable at detecting both hard and soft faults than the  na\"{i}ve-approach, especially when the network outage is high.   %show reliable performance of the proposed methods 
Furthermore, we test the methods on real-world data from a smart metering deployment. The methods, in particular the clustering method, are shown to be applicable and useful in a real-world scenario. %\rbs{re-evaluate abstract before submission: more results}
\end{abstract}

\begin{IEEEkeywords}
Internet of Things, IoT, Network Monitoring, Quality of Service, QoS, Machine Learning, Unevenly Spaced Time Series, Lomb-Scargle,
\end{IEEEkeywords}

% \fi
\IEEEpeerreviewmaketitle

\section{Introduction}

%\IEEEPARstart{T}{he} Internet of Things $\rightarrow$ massive data. Enabled by IoT networking protocols. Network data potential: Analysis of networks, monitoring of IoT deployments, scalability analysis and provisioning.

%%%%% - General intro to massive IoT systems, the fact that the focus has been on developing techniques and protocols. However, once system is installed and operates, we need to have a continous insight in its performance, link status, etc. Put a figure with a general IoT architecture (two BSs connected to a cloud, devices connected to the BS). This motivates the functionality of a tool that helps in analysis which we will refer to as a IoT Dashboard. 

\begin{figure}[tb]
    \centering
    \includegraphics[width=.95\columnwidth,trim={0cm 0cm 0 0},clip]{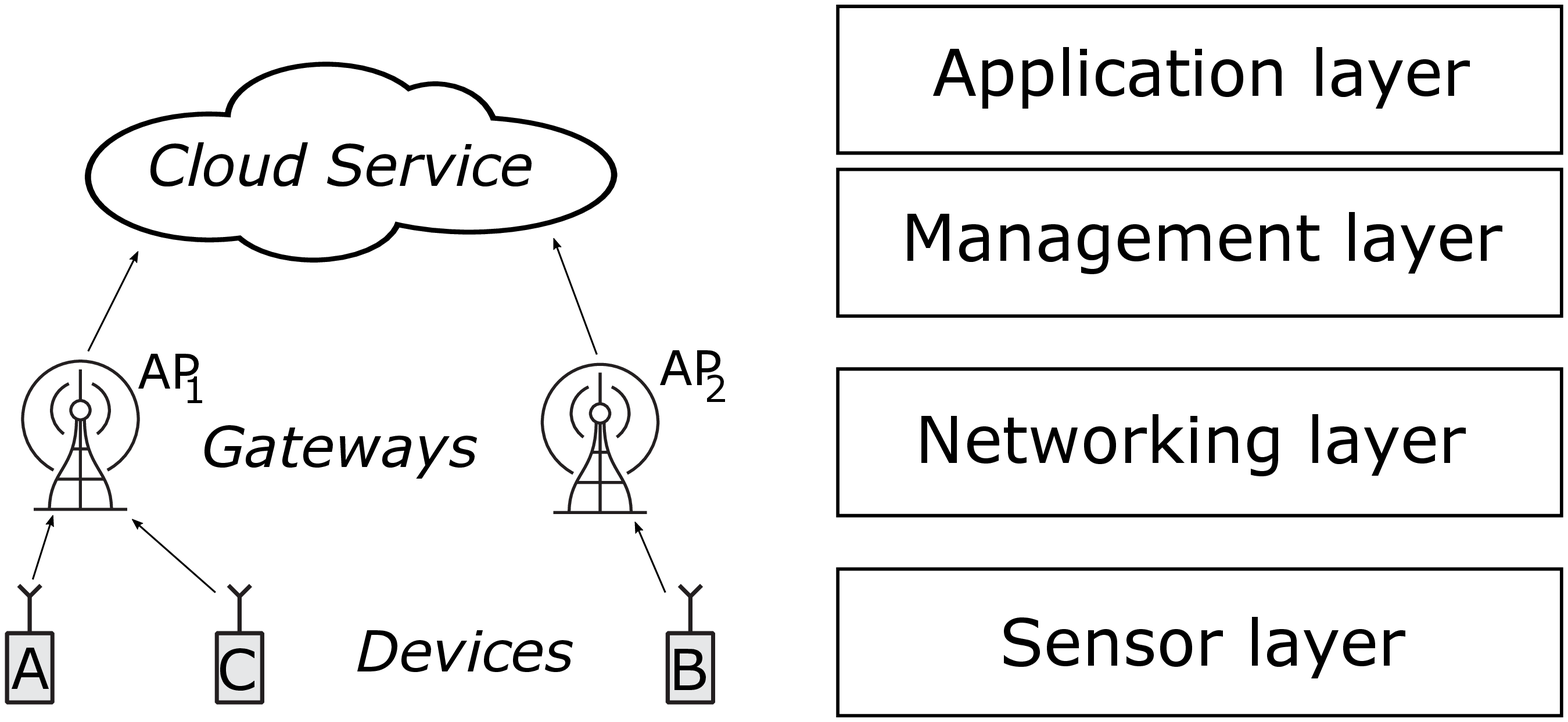}
    \caption{IoT deployment topology and architecture. A gateway could be, for example, a cellular towers, a LPWAN access point, or a satellite. The network topology between devices and gateways may be a mesh or a star-topology. The IoT deployment is connected to a virtual IoT server in the cloud, which manages the network and makes application features available to end users. The different tiers of the network are directly sensing different information about the network. Information regarding link quality of connected devices is available to the APs, but only to the management tier if this information is relayed.%\PP{This caption should reflect the fact that different information is available at different points in the network. You are discussing the public/private thing further in the text, this needs to be concisely reflected here.}
    }
    \label{fig:deploymentArch}
\end{figure}

\IEEEPARstart{I}{ot} deployments can provide a large variety of services capable of cyber-physical interactions through sensors, actuators and data analysis by utilizing fog or cloud computing. 
Such deployments can be cyber-physical systems, consisting of devices that exchange messages with servers through networks, as depicted in Fig.~\ref{fig:deploymentArch}, which also shows the common IoT architecture~\cite{iotsurvey}. In such deployments the traffic generated by sensors is intermittently filtered by the network before being received by the IoT server. This intermittency is caused random network effects, such as medium access delays, outage in the network and queuing of transmissions.

%Such deployments are cyber-physical systems, not only due to interactions with the real-world at the devices layer, but also due to the interactions between the Network layer and the real world. 
% \textcolor{red}{Is it really necessary to introduce all of these new acronyms? These are some that you have come up with, right? I find it a bit distracting.}
% Such deployments can provide a large variety of services capable of cyber-physical interactions through sensors, actuators and data analysis by utilizing fog or cloud computing. 

During the past decade, techniques and standards have been developed to provide adequate networking features and improve the Quality of Service (QoS) for the vast number of IoT use cases. An overview of the architecture of IoT services and enabling technologies and protocols is given in~\cite{iotsurvey}. Specifically, in the context of wireless communications, the term IoT usually refers to \textit{massive machine-type communication} (mMTC), one of the three connectivity types in 5G~\cite{NR5g}. Here, a number of pre-5G IoT systems for massive connectivity have been developed and are currently being deployed. These include the low power wide area networks (LPWANs): SigFox, LoRaWAN, NB-IoT and LTE-M~\cite{lpwans,nbiot,ltem}. 

A central problem of wireless IoT connectivity is monitoring and status detection for a massive number of connected devices, which provides insights into the status of the links and devices and can potentially lead to corrective actions ~\cite{iotsurvey,netmonsurvey,netmonsurvey2}. Network monitoring for wireless sensor networks (WSN) has been researched for decades. A comprehensive survey of network monitoring in wireless sensor networks (WSNs) can be found in~\cite{netmonsurvey} and~\cite{netmonsurvey2}. Arguments for the importance of network monitoring is given by these surveys; network monitoring can enhance data reliability, bandwidth utilization, and the lifetime of the WSN due to the opportunity to identify faulty devices and hence better utilize constrained resources. Multiple monitoring methods can be combined through frameworks, such as  fuzzy logic~\cite{fuzzyframe} to optimize decision making for fault tolerance. In general, the fault detection methods for WSNs assume a PAN mesh topology such as 6LoWPAN or ZigBee. For example, PAD, a passive monitoring method relying on inference based on routing changes is presented in~\cite{pasDiag}. 
Nevertheless, LPWANs are one-hop star-topology networks and additionally, PAD introduces a few bytes of overhead to transmissions, which is a major drawback for energy-constrained devices and networks supporting massive numbers of devices.
Notably, the fault detection methods of~\cite{netmonbayes} and~\cite{netmonar} rely on statistical inference based on the timing of incoming traffic to detect faults in the network. These methods may readily be modified for and applied to LPWANs since they do not rely on topology changes in a mesh network, but they have the drawback of requiring fault-less training data in order to recognise healthy behavior and the detected errors are only quantifiable at a low resolution, ie. "no errors, some errors or many errors". %\PP{You need to clearly explain what do you mean by a resolution here and why it is important, since a couple of lines below it is advertised that the proposed solutions have a good resolution.}

% but the prevalence here is to assume a mesh network topology and infer network faults by observing topology changes or introducing transmission overhead such as active probing. These assumptions are not valid for all IoT deployments, especially not in the context of LPWANs, which are star-topology networks for resource-constrained devices and long-range communication.

 %\rbs{Monitoring LPWANs and cellular networks are in particular relevant as these technologies are currently being deployed. RBS: reiteration, should I delete?}

\begin{figure}[tb]
    \centering
    \includegraphics[width=.95\columnwidth,trim={0 1.5cm 0 0},clip]{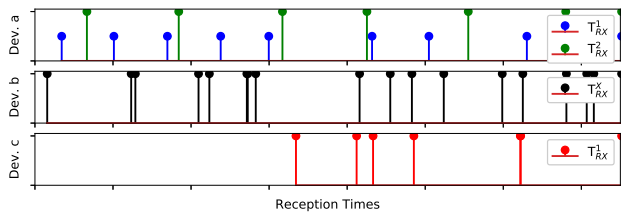}
    \caption{Devices generate traffic depending on their applications. Here the traffic patterns are depicted for three types of devices; Dev. a) runs two applications that generate quasi-periodic traffic. The received traffic streams are shown for both labeled and unlabeled traffic. Dev. b) runs three quasi-periodic applications, but does not label the traffic. %Dev. c) runs a single event-driven applications, in the depicted case the inter-arrival time is exponential.
    }
    \label{fig:devtraffic}
\end{figure}

In this paper, we make no assumption about the topology of the network and instead we assume a quasi-periodic\footnote{Quasi-periodic applications generate transmissions at a constant, or near constant, inter-arrival time. See Sec.~\ref{sec:traffic} for more details.} traffic model as that depicted in Fig.~\ref{fig:devtraffic} for device A and B, which is common for IoT applications. This approach allows us to evaluate the state and link performance of individual devices passively, which is important for identifying poor performers and malfunctioning devices in massive IoT deployments. In our model devices can run 'thin' clients (a single application) or 'thick' clients (multiple applications). We have observed the latter behaviour in a data-set from a LoRaWAN deployment, where mains-powered sensors and actuators were used to control and manage street lights in rural towns. %Furthermore, known 'thin' clients exhibiting 'thick' client behaviour may indicate a byzantine device or network.
We propose parametric machine learning methods for high resolution, centralised and passive fault detection in arbitrary IoT deployments. The methods use temporal correlations in observed traffic to parameterize quasi-periodic applications. The modelled applications are used to infer the state and the QoS of individual devices. Interestingly, the methodological basis for this work has been drawn from research in astronomy, where unevenly sampled time series are common. Astronomers have developed techniques for analyzing such series including phase-folding~\cite{pdm} and a variant of the classical Fourier periodogram that is generalised for uneven time-series \cite{Lomb1976,Scargle1983,Baluev2008,VanderPlas_2018}.

{The paper is structured as follows: We introduce our traffic model, the traffic meta-data that can be expected to be available for analysis, and targeted network KPIs in Sec.~\ref{sec:dash}. In Sec.~\ref{sec:lab} we analyze the sub-problem of parametric regression for traffic that is labeled by its parent application. The classification problem of labeling traffic when the traffic parameters are known, but the parent applications are unknown, is examined in Sec.~\ref{sec:labeling}. In Sec.~\ref{sec:unlab} we treat the clustering problem that arises when no a priori information is given. Performance results for the algorithms presented throughout can be found in Sec.~\ref{sec:results} and the results of using the fault-detection algorithms on a real-world smart metering deployment can be found in Sec.~\ref{sec:realdat}. Sec.~\ref{sec:conc} contains concluding remarks. {Table \ref{tab:notes} on page \pageref{tab:notes} provides a list of symbols and mathematical notations.}}

\section{System Model and Key Performance Indicators} \label{sec:dash}

% Let an IoT deployment be a cyper-physical system consisting of sensory and actuating devices, a computing entity in the form of a cloud or fog server and one or more networks, which provide connectivity between the former. 
% The network can be seen as a filter, which acts upon any transmissions passing through it, introducing jitter in the form of delays and transmission loss. %In this case the server is acting as a data sink in the deployment in order to provide a service. Another case is that the server act as a central control hub, which generates and transmits commands to the devices.
% \rbs{introduce network monitoring concept - in short~\cite{netmonsurvey,netmon2019}}
% IoT Networks focus on providing connectivity for a massive number of sporadically transmitting devices %contrary to keeping a duplex channel open with every single device \rbs{[ref?]}. 

Consider an IoT deployment like that of  Fig.~\ref{fig:deploymentArch}~\&~Fig.~\ref{fig:devtraffic} where sporadically transmitting devices are  connected to a server by an arbitrary networking technology, for example an LPWAN, or a mesh network. Here, a virtual server running in the cloud acts as a centralised management layer for network monitoring. The wireless network acts as a filter upon transmitted data introducing intermittency in the form of outage and delays in the received data. In this section we generalise this model to any number of gateways and devices running any number of applications. First, we present the meta-data available in a passive monitoring scenario, define a traffic model for quasi-periodic IoT devices, a set of key performance indicators (KPIs) and a  na\"{i}ve method for evaluating the KPIs. Lastly, we examine how meta-data changes the nature of the machine learning problem. 
% In this section we define a traffic model for IoT devices and %which we base our fault detection methods on and 
% present the meta-data available in a passive monitoring scenario as well as the key performance indicators (KPIs) common to devices and networks.

%Under such conditions the computing entity will experience the transmissions received from sensory devices in their filtered form. In this work we focus on performance monitoring of devices and network from this 'server-side' perspective.
% In this work we focus on the case where the server acts as a data sink, but the methods developed here may also be used locally in each device for analysis of the network performance provided that the traffic model of the server commands fits with the 'quasi-periodic' traffic model for devices that we will now define.

\subsection{Available Traffic Meta-Data}

%\PP{Try to relate this part to Figure 1 where you can explicitly denote the points at which a specific metadata is available under a certain assumption of private/public access.}
%The meta-data that is available for analysis depends on the network type and technology. 
In Fig. \ref{fig:deploymentArch} the devices A and C transmit data through AP$_1$ while device B transmits data through AP$_2$. The data that is available for analysis at the server depends on how much meta-data AP$_1$ and AP$_2$ relay to the server in addition to how much meta-data is included in the transmissions.
When a network is licensed from a network provider then it is considered to be a public network~\cite{6839076}. A network that is closed to the public, privately owned or purpose-built specifically for an IoT deployment, is considered a private network. The same network can be considered private to its owner and public to licencees.
In private networks link level metrics, such as %RSSI, LQI, SNR, CSI, 
received signal strength indicator, link quality indicator, signal to noise ratio, channel state information, 
modulation and coding rate, are available whereäs they are not necessarily available in public networks. Some network technologies and protocols have specific meta-data built in to the protocol, for example GPS coordinates in SigFox. Such metrics can not always be expected to be available in public networks.

We define a minimal set of metrics that are available in any type of network, \{Network ID, Device ID, Reception Timestamp, Payload size\}. Device and Network identifiers are a necessary part of a useful transmission. The transmission size and reception time can likewise be found for any transmission. This ensures that the network monitoring is applicable both for network operators, who licence their networks to IoT deployments, but do not have in-depth knowledge of the IoT deployment or access to transmitted data and IoT vendors and operators, who do not have insight into the networking components or access to link-level meta-data.

In addition to this minimal set of metrics, devices may include the ID of their parent application as meta-data within the transmission. This is the behavior of device A in Fig. \ref{fig:devtraffic} in contrast to device C, which does not label its traffic. The implication of labeling is that we may pose the fault detection problem as a regression problem instead of a clustering problem. We discuss this implication further in Sec. \ref{sec:priori} after defining our traffic model and KPIs.

% We define a minimum set of available metrics for generality. This set is composed of: 
% \begin{itemize}
%     \item Service ID
%     \item Network ID
%     \item Device ID
%     \item Reception Timestamp
%     \item Payload size
% \end{itemize}

% As mentioned in the introduction, we wish to use this data for monitoring, scalability analysis and provisioning. These are the three functionalities of the dashboard entity, which we will elaborate on in Sec.~\ref{sec:dashfunc}.

\subsection{Traffic model} \label{sec:traffic}
The traffic generated by a device is the composite of the traffic generated by all the applications running on the device. We denote the number of apps running on a device by $I_{\mathrm{apps}}$. Traffic of one app may influence the jitter of traffic of another app due to queuing of transmissions in the device. 
\begin{align} \label{eq:peridiccomposite}
    T^{x}_{\mathrm{RX}} = T^1_{\mathrm{RX}} \cup T^2_{\mathrm{RX}} \cup ... \cup T^{(i)}_{\mathrm{RX}} \cup ...  \cup T^{I_{\mathrm{apps}}}_{\mathrm{RX}}
\end{align}
Where $T^i_{\mathrm{RX}} =$ \{$T^i_1, T^i_2, ..., T^i_m, ... T^i_{M-1}, T^i_M$\} denotes the received transmissions from application $i$. 
\\\noindent We define Quasi-periodic applications:

% \begin{figure}[tb]
%     \centering
%     \includegraphics[width=.95\columnwidth,trim={.7cm 1cm 0 0},clip]{traffic_example.eps}
%     \caption{Example of labeled and unlabeled quasi-periodic traffic from a 'thick' device running three applications. In this example $\mathrm{p_{o}} = 0.2$ and $0<J^i_{m^i}<\alpha^i/2$ where $J^i_{m^i} \sim \exp(0.25)$.}
%     \label{fig:traffic}
% \end{figure}

% \subsubsection{Quasi-periodic traffic}
These are applications, which send periodic reports, but where the received traffic is intermittent due to queuing and filtering by the network.
Common IoT use cases such as gas-, water- and electric smart metering, smart agriculture and smart environment \cite{3gppMAR,6740844}, are considered by 3GPP to be quasi-periodic~\cite{3gppMAR}.
%The Mobile Autonomous Reporting (MAR) periodic reports considered in~\cite{3gppMAR} is quasi-periodic. Expected MAR-p use cases include gas-, water- and electric smart metering, smart agriculture and smart environment \cite{3gppMAR,6740844}, which are all also use-cases for LPWANs.

The $i$th quasi-periodic app generates transmissions at approximately constant intervals, such that the reception times of transmissions from app $i$ can be described by:
\begin{align} \label{eq:peridic}
    T^i_{{m^i}} = \beta^i + \alpha^i\cdot (m^i+o^i_{m^i}) + J^i_{m^i}
    %\\
    % T_i(k) = \beta_i + \alpha_i\cdot (k+o_i(k)) + J_i(k)
\end{align}
where $\beta^i$ is a time offset, $\alpha^i$ is the inter-arrival time of app $i$, $J^i_m$ is a random delay introduced by the network (jitter), $m^i$ is the index of the received packets while $o^i_{m^i}$ is the cumulative number of transmissions that were not received until observation $m^i$. Let $n^i$ be the index of the transmissions such that $n^i = m^i + o^i_{m^i}$.
Then we know that $T^i_{n-1}  \cong T^i_{n} \cong T^i_{n+1} \pmod{\alpha^i}$ for the quasi periodic app $i$.

\subsection{Key Performance Indicators} \label{sec:dashfunc}

We wish to monitor % the health of the network and 
the link quality and status of individual devices in IoT networks. Based on the available meta-data we choose to monitor network outage and online/offline status, which are common KPIs in wireless network performance modelling. % and analogous to the mean-time between failures and downtime in process management literature.
In this subsection we define each of these KPIs.

\subsubsection{Offline detection}
% The state of devices can be detected either from the throughput given a known outage probability or from the number of unobserved expected transmissions.
An app or a device is considered to be offline if it stops generating transmissions.
An intuitive classifier for offline status is detecting whether $k$ consecutive expected transmissions have been missed. This can be done at the application level to classify applications as offline or at the device level, to classify an entire device as offline. Offline applications are not expected to generate transmissions and so they do not count towards the calculated outage. We define a classifier for offline entities, $C_\text{Off}$.
\begin{align}
    C_\text{Off}: o^i(t) \geq k
\end{align}
where $o^i(t) = \bigg{\lfloor}\dfrac{t-T^i_{m^i}}{\alpha^i}\bigg{\rfloor}$ computes the expected number of transmissions at time $t$ since the last reception at $T^i_{m^i}$ if the quasi-periodic app $i$ was online.

% We compute $k$ as a function our expected network outage $\mathrm{p_{o}}$ and arrival rate 1$/\alpha$. 
% \begin{align}
%     k(\mathrm{p_{o}},\alpha) = \max \bigg(\dfrac{\log_{10}(\alpha)}{\log_{10}(\mathrm{p_{o}})}, 3\bigg)
% \end{align}

\subsubsection{Outage probability}
%Outage introduced in the network can be estimated from the throughput if a hypothesis for the true throughput is put forward and $P_\text{OffOn}$ and $P_\text{OnOff}$ are known or assumed. Equal values of $P_\text{OffOn}$ and $P_\text{OnOff}$ correspond to a network, which is in a wide sense stationairy (wss) state. In the wss case we may estimate $P_{out}$ as
%
%\begin{align} \label{eq:outthr}
%    P_{Out} = \dfrac{T_\text{E}(t,\tau_\text{W})-T_\text{O}(t,\tau_\text{W})}{T_\text{E}(t,\tau_\text{W})}
%\end{align}
%
%%  The accuracy of this estimate is constrained by arrival rate, the number of devices considered and window size. 
%
%
%
%Where $T_E$ is the expected throughput and $T_O$ is the observed throughput. 
%
%Eq. \eqref{eq:outthr} can be rewritten in a more generic form in terms of the number of expected packets and observed packets over a window $T_W$ at time $t$ are denoted by $n_E(t,T_W)$ and $n_O(t,T_W)$, respectively.
We define the outage probability as the ratio of the number of packets lost to the number of transmitted packets over a window $\tau_w$ at time $t$.

\begin{align}
    {p_\mathrm{o}^*}= \sum_{i=1}^{I_{\mathrm{apps}}}o_{m^i(t,\tau_w)} \Bigg{/} \sum_{i=1}^{I_{\mathrm{apps}}}n^i(t,\tau_w)
\end{align}
 where the number of transmitted and received packets for app $i$ from time $t-\tau_w$ to $t$ are denoted by $n^i(t,\tau_w)$ and $m^i(t,\tau_w)$, respectively.
As $n^i(t)$ approaches $\infty$ the observed outage, ${p_\mathrm{o}^*}$, approaches the network outage probability, ${p_\mathrm{o}}$, if ${p_\mathrm{o}}$ is stationary over observation period.

$o_{m^i(t)}$ and $n^i(t)$ are latent variables from the IoT server's perspective. Then the goal of fault detection can be posed as the problem of estimating these latent variables correctly.

% \begin{enumerate}
%     \item Choose a traffic model according to domain knowledge
%     \item Fit model parameters to traced traffic
%     \item Forecast expected traffic using the parameterized model
%     \item Estimate outage based on expected and observed traffic
% \end{enumerate}

\subsection{Na\"{i}ve monitoring method} \label{sec:naive}
Here, we introduce a  na\"{i}ve monitoring method, which we will compare other methods to. The approach is straightforward; Denote the number of transmissions received within $\tau_w$ at time $t$, $m_{\mathrm{obs}}(t,\tau_w)$, and let $m_{\mathrm{max}}$ denote the maximum value of $m_{\mathrm{obs}}(t,\tau_w)$ for $t < t_{now}$. Then we have \eqref{eq:naiveout}.
\begin{align} \label{eq:naiveout}
    {p_\mathrm{o}^*}(t,\tau_w) = \dfrac{m_{\mathrm{max}}-m_{\mathrm{obs}}(t,\tau_w)}{m_{\mathrm{max}}}
\end{align}

Devices where ${p_\mathrm{o}^*}(t,\tau_w) > \epsilon$ are classified as offline.

\subsection{A priori knowledge and labeled traffic} \label{sec:priori}

% In this work we aim to estimate the outage of devices, which in general, can be done by parameterizing a traffic model, that is chosen based on domain knowledge, and hypothesize what the number of expected packets is, $n_H$. If our traffic model and parameters are good fits we have that $n_T \approx n_H$.

We may have a~priori knowledge about online devices and applications, and we may receive meta-data identifying the app source of traffic. This changes the nature of the fault monitoring problem as depicted in Table \ref{tab:prob}. 
% \begin{itemize}
%     \item In the case where we know the parameters of the traffic model and transmissions are labeled (or all clients are 'thin'), we can calculate the KPIs straightforward. 
%     \item In case the traffic model is unknown, but traffic is labeled we must perform a regression of the traffic parameters in order to calculate the KPIs, which is treated in Sec.~\ref{sec:lab}. 
%     \item In Sec.~\ref{sec:labeling} we examine the case that traffic parameters are known, but the traffic is unlabeled (and from a 'thick' client) we must classify received packets as belonging to one app or another to evaluate the KPIs.
%     \item Finally, we tackle the problem of clustering when the traffic parameters are unknown and traffic is unlabelled in Sec.~\ref{sec:unlab}.
% \end{itemize}
In the case where we know the parameters of the traffic model and transmissions are labeled (or all clients are 'thin') it is straightforward to calculate the KPIs. In case the traffic model is unknown, but traffic is labeled we must perform a regression of the traffic parameters in order to calculate the KPIs, which is treated in Sec.~\ref{sec:lab}. In Sec.~\ref{sec:labeling} we examine the case that traffic parameters are known, but the traffic is unlabeled (and from a 'thick' client) we must classify received packets as belonging to one app or another to evaluate the KPIs. Finally, we tackle the problem of clustering when the traffic parameters are unknown and the traffic is unlabelled in Sec.~\ref{sec:unlab}.

\begin{table}[tb]
\resizebox{\columnwidth}{!}{
\begin{tabular}{|l|l|l|l|}
\hline
\multicolumn{2}{|l|}{Determining problem 
%\textbf{\begin{tabular}{l}\\ Diagram for determining \\ the problem type\end{tabular}}
} & \multicolumn{2}{l|}{%\begin{tabular}{l}The number of apps of each device and \\ the parameters of each app are known\end{tabular}
\textit{Known applications} } \\ \multicolumn{2}{|l|}{nature and applicable } & \multicolumn{2}{l|}{\textit{\& traffic parameters}} \\
\cline{3-4} 
\multicolumn{2}{|l|}{ methodology} & \textit{Yes} & \textit{No} \\ \hline
\multirow{2}{*}{%\begin{tabular}{l} The transmission is labeled with \\ the ID of the parent application \\ \end{tabular}
\textit{Labelled traffic}
} & \textit{Yes} & KPI calculation 
& Regression \\ \cline{2-4} 
 & \textit{No} & Classification & Clustering \\ \hline
\end{tabular}
}
\vspace{.01cm}
\caption{Diagram of the classes of the fault detection problem.}
\label{tab:prob}
\end{table}

\section{Regression and Classification} \label{sec:sup}

\subsection{Regression} \label{sec:lab}
In case transmissions are labeled by their parent application, the composite stream of received packets from all applications can easily be sorted by application such that the stream of any periodic application is on the form of \eqref{eq:peridic}. We wish to learn $\alpha^i$ and $o^i_{m^i}$ given a set of reception times, $T^i_{\mathrm{RX}} =$ \{$T^i_1, T^i_2, ... T^i_m ... T^i_{M-1}, T^i_M$\}. Phase-folding methods like phase dispersion minimization~\cite{pdm} could solve this problem in a brute-force manner by estimating the fit of all potential $\alpha^i$, but the associated computational overhead is undesired for massive online network monitoring.%. However, such techniques require estimating the fit of every candidate period, which can be a very large set, making phase-folding methods a brute-force approach that is undesirable for online network monitoring.

Instead, we propose Normalised Harmonics Mean (NHM), which is an online method we developed 
for finding an estimate $\alpha^{i*}$ of $\alpha^i$ in a set $T^i_{\mathrm{RX}}$ with a relatively low amount of computational effort. 
%, then we make a new guess for $\alpha^i$ and keep reiterating this procedure until $\alpha^{i*}$ converges. 

Consider a MMSE of the distance between a reception time and transmission time, $\min(|T^i_{{n^i_{m^i}}}-T^i_{{m^i}}|^2)$ where $n_m^i = m^i+o_{m^i}$. Here $o_{m^i}$ and $n_{m^i}$ are unknown to the receiver, but we know that the generating function for $T^i_{n_{m^i}}$ is periodic. We may solve the problem by brute-force using least-squares, however this requires finding $n_{m^i}$, which minimizes the MSE for every arrival ${m^i}$ for every proposed $\alpha^i$, which makes this solution computationally expensive and the accuracy depends on the $\alpha^i$ grid chosen for the analysis. This is similar to the brute-force approach of phase dispersion minimization~\cite{pdm}.

%We may find $n_{m^i}$ by trying all $n_{m^i}$ from ${m^i}$ to $o_{m^i}+1$ and choosing $n_{m^i}$ to be the value, which minimizes the MSE. 
We propose to use a periodic function, specifically the cosine, to describe the problem. Then we have \eqref{eq:txnhm} at the transmitter side and \eqref{eq:rxnhm} at the receiver side.
\begin{align} 
    &\cos\Bigg(\dfrac{T^i_{{n_m^i}}-T^i_{{n_m^i-1}}}{\alpha^i}\cdot 2\pi\Bigg) = 1 \label{eq:txnhm} \\ 
    &\cos\Bigg(\dfrac{T^i_{{m^i}}-T^i_{{m^i-1}}}{\alpha^i}\cdot 2\pi\Bigg) > 0 \quad , \quad 0<J^i_{m^i}< \dfrac{\alpha^i}{2} \label{eq:rxnhm}
\end{align}

So we need to solve for $\alpha^i$ that maximizes \eqref{eq:rxnhm}, which may be rephrased as trying to get ${(T^i_{{m^i}}-T^i_{{m^i-1}})}/{\alpha^i}$ to be as close to an integer as possible. %Once $\alpha^i$ is found $\beta^i$ can be calculated by maximizing ${(T^i_{{m^i}}-\beta^i)}/{\alpha^i}$.

NHM is based on this observation and attempts to find $\alpha^i$ in a gradient-descent manner, by searching for the set of best fitting integers to normalise the distances between elements in $T^i_{\mathrm{RX}}$.
% In NHM an intial hypothesis for $\alpha^i$ is made and the true $\alpha^i$ is found in a gradient-descent manner, by normalising the distances between elements in $T^i_{\mathrm{RX}}$.
The step-wise procedure of NHM is described by Alg.~\ref{alg:nhm}.% as depicted in Fig \ref{fig:labelHarmonic}.
% The step-wise procedure of NHM follows:% as depicted in Fig \ref{fig:labelHarmonic}.

% \begin{enumerate}
%     \item Make a preliminary hypothesis for $\alpha^{i*}$.
%     \item Estimate $\alpha^{i*}$.
%     \begin{enumerate}
%         \item Estimate $\eta^i_{m^i}$ for $\alpha^{i*}$.
%         \item Make new hypothesis for $\alpha^{i*}$.
%         \item Reiterate step 2 until $\alpha^{i*}$ converges.
%     \end{enumerate}
%     \item Calculate KPIs based on app parameters.
% \end{enumerate}

\begin{algorithm}[b]
\KwData{$dT^i_{\mathrm{RX}}$}
\KwResult{$\alpha^{i*}$}
    Make a preliminary hypothesis for $\alpha^{i*}$.\\
    \While{$\alpha^{i*}$ \text{has not converged}}
    {
        Estimate $\eta^i_{m^i}$ for $\alpha^{i*}$.\\
        Update the hypothesis for $\alpha^{i*}$ based on $\eta^i_{m^i}$.
    }
    Calculate KPIs based on app parameters.
    \caption{Normalised Harmonics Mean (NHM)}\label{alg:nhm}
\end{algorithm}

\begin{figure}[tb]
    \centering
    \includegraphics[width=1\columnwidth,trim={0 .4cm 0 1.2cm},clip]{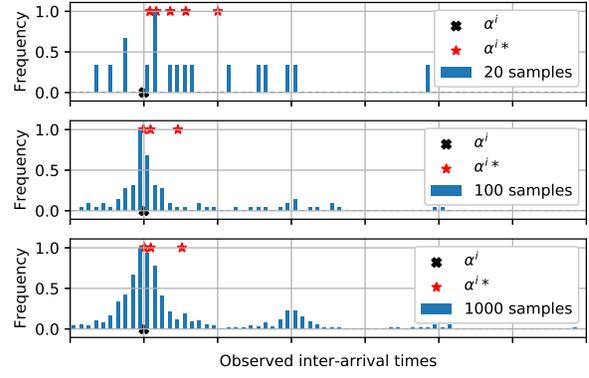}
    \caption{Histogram of $dT^i_{\mathrm{RX}}$ where $J^i_m~\sim~\exp(.25)\cdot{\alpha^i}{/2}$ and ${p_\mathrm{o}} = 0.2$. The red crosses indicate the iterations of $\alpha^{i*}$ in NHM, which can be seen to approach $\alpha^{i}$, even when few samples are available. In this example, the initial hypothesis of $\alpha^{i*}$ was based on the mean of $dT^i_{\mathrm{RX}}$.}
    \label{fig:labelHarmonic}
\end{figure}

Let the set $dT^i_{\mathrm{RX}}$ denote the set of the distances between neighboring elements $T^i_{\mathrm{RX}}$. The preliminary hypothesis for $\alpha^i$ %could either be $\alpha^{i*} = \min(dT^i_{\mathrm{RX}})$ or $\alpha^{i*} = \mathrm{mean}(dT^i_{\mathrm{RX}})$.
could be $\alpha^{i*} = \mathrm{mean}(dT^i_{\mathrm{RX}})$.
% A preliminary hypothesis for $\alpha^*$ is required. Two cases for selection come to mind: Letting $\alpha^*$ be the minimum observed time between packets and letting $\alpha^*$ be the mean  observed time between packets. 
% The former preliminary estimate is very robust to jitter and outage, but highly sensitive to mislabeled transmissions. The other option is robust to mislabeled packets. %However, the latter option makes NHM converge faster, which minimizes computation time and so it is preferred here.

% \begin{align} \label{eq:labelhypo1}
% \alpha^* = \min(dT^i_{\mathrm{RX}})
% \end{align}

Outage creates harmonic contributions of orders higher than 1, $o^i_{m^i} > 0$, in the set $dT^i_{\mathrm{RX}}$ as depicted in Fig. \ref{fig:labelHarmonic}. In the next step we seek to normalize these harmonics by estimating the series of latent variables $\eta^i_{m^i}$ given by \eqref{eq:labeln}. We do this under the condition that $||J|| < \dfrac{\alpha^i}{4}$ (or equivalently $0 \leq J < \dfrac{\alpha^i}{2}$) such that aliasing is avoided, $n^i_{m^i} \leq n^i_{m^i+1}+1$, which in practice means that a transmission is never received earlier than a previous transmission.% under the hypothesis of $\alpha^* = \mathrm{mean}(dT^i_{\mathrm{RX}})$. 
\begin{align} \label{eq:labeln}
\eta^i_{m^i} = \bigg{\lfloor} \dfrac{T^i_{m^i}-T^i_{m^i-1}}{\alpha^{i*}} \bigg{\rceil}
\end{align}
where $n^i_{m^i} = \sum_{k=1}^{{m^i}} \eta^i_{k} \quad \Leftrightarrow \quad \eta^i_{m} = n^i_{m}-n^i_{m-1}$. 

The initial hypothesis for $\alpha^{i*}$ is unlikely to be correct, but it serves to provide an initial estimate of the latent series $\eta^i_{m^i}$, which we can now use to estimate $\alpha^i$. %Using $\alpha^{i*} = \min(dT^i_{\mathrm{RX}})$ as a preliminary estimate is very robust to jitter and outage, since neither changes the value of our first estimate of the latent series unless $||J|| \geq \dfrac{\alpha^i}{4}$. However, it is highly sensitive to mislabelled transmissions, which can seriously change our initial estimate of the latent series. In the case mislabelled packets are present, we are better of using $\alpha^{i*} = \mathrm{mean}(dT^i_{\mathrm{RX}})$ as a preliminary estimate, trading robustness to jitter and outage for robustness to mislabelled transmissions.
Given the hypothesis that $\alpha^i$ is the mean of a normalised version of $dT^i_{\mathrm{RX}}$ we have~\eqref{eq:labelIAupdate}.
\begin{align} \label{eq:labelIAupdate}
\alpha^{i*} = \dfrac{1}{M^i} \sum_{m^i=2}^{M^i}{\bigg{(}\dfrac{T^i_{m^i}-T^i_{m^i-1}}{\eta^i_{m^i}}\bigg{)}}
\end{align}

Iterative updates of $\eta^i_{m^i}$ and $\alpha^{i*}$ Eq. \eqref{eq:labelIAupdate} will gradually go towards $\alpha^i$ as depicted in Fig.~\ref{fig:labelHarmonic}.
The required number of iterations for convergence depends on the accuracy of the initial estimate of $\alpha^i$, which is dependent on ${p_\mathrm{o}}$ and the number of samples. After the algorithm convergences the estimated number of transmissions lost between successive receptions is given by the latent variables $o^i_{m^i} = n^i_{m^i}-n^i_{m^i-1}$. Furthermore $\beta^i$ can be found by linear regression of \eqref{eq:peridic} as both $\alpha^i$ and $n^i_{m^i}$ have been estimated, but is not necessary for computing the KPIs in our case.
%The more samples that are used, the faster the convergence is.
% \begin{figure}[tb]

        % \caption{a) Histogram of $IA$ observations and $IA^*$ iterations in NHM of 100000 samples of $T_m$ where $IA = 100$ and $\mathrm{p_{o}} = .1$. The NHM method converges on the mean of the IAs, while attempting to normalize n'th order harmonics to the 1st-order distribution. b) Mean error of NHM over 1000 realisations of $N_{m+o}$ = {3, 6, 10, 100} as a function of the outage in $N_{m+o}$. The $IA$ and J are randomly selected for each run, but limited by $J < IA/4$ }
% \end{figure}

% An algorithmic description can be found in Algo.~\ref{alg:labeled}. 
NHM can be run in an online manner by saving $dT^i_{\mathrm{RX}}$ and $T^i_{M+1}$ and updating both upon reception of a new transmission for app i before repeating step 2 starting from the previous $\alpha^{i*}$. %The iterative steps need not be recalculated over all sampled data unless $o^i_{M+1} \neq 0$. 

%Here, we have estimated $\alpha^i$ under the assumption that the jitter will be zero-mean. Given other distributions of J, the mean-offset of the jitter should be accounted for. NHM does however suffice in our intended application where $||J||<<\alpha$. We note that the accuracy of this type NHM has been found to be good also for exponentially and normally distributed jitter. 
% However, NHM relies on prior labeling of packets, which may not always be available. 

% \begin{figure}[tb]
%     \centering
%     \removelatexerror
%     \begin{algorithm}[H]
%         \KwData{$T$ and $\epsilon$}
%         \KwResult{$\alpha^*$}
%         M = length($T$) \;
%         \tcc{Make initial estimate of $\alpha$*}
%         $T_{min}$ = $T_{1}$ \;
%         \For{m=2:M}{
%         $T_{diff_{m-1}} = T_{m}-T_{{m-1}}$ \;
%         \If{$T_{diff_{m-1}} < T_{min}$}
%         {$T_{min} = T_{diff_{m-1}}$ \;}
%         }
        
%         $\alpha^* = T_{min}$ \;
%         \tcc{Reestimate $\alpha$* iterately until converge}
%         $\alpha^*_{prev} = \alpha^*+2 \cdot \epsilon$ \;
%         \While{$|\alpha^* - \alpha^*_{prev}| > \epsilon$}{
%         \For{m=1:M}{
%         $n_m$ = $\lfloor T_{m}$ / $\alpha^* \rceil$ \;
%         $S_{m} = \dfrac{T_{m}}{n_m}$\;
%         }
%         $\alpha^*_{prev} = \alpha^*$ \;
%         $\alpha^* = \dfrac{1}{M}\cdot \sum(S_{m})$ \;
%         }
%         \caption{Normalised Harmonic Mean: NHM()}
%         \label{alg:labeled}
%     \end{algorithm}
% \end{figure}

\subsection{Classification} \label{sec:labeling}
In case the parameters $\alpha^i$ and $\beta^i$ are known for all applications $i\in I$, but transmissions are not labeled by their parent application, we need to classify which application a new reception belongs to. 
Let $f(T^i_{\mathrm{RX}},\alpha^i,\beta^i)$ be the transmission generating function, then we may propose a periodic likelihood function $g(T^i_{\mathrm{RX}},\alpha^i,\beta^i)$, such that 
\begin{align}
    f(T^i_{\mathrm{RX}},\alpha^i,\beta^i) \propto g(T^i_{\mathrm{RX}},\alpha^i,\beta^i) = \dfrac{\cos\bigg(\dfrac{T^i_{\mathrm{RX}}-\beta^i}{\alpha^i}\cdot 2\pi\bigg)+1}{2}
\end{align}
Then $i$ is found by $\max_{i∈\{1:I_{\mathrm{apps}}\}}(g(T^i_{\mathrm{RX}},\alpha^i, \beta^i))$. Here, we chose a cosine function as a likelihood function and added 1 before normalizing to keep the probability between 0 and 1. In practice the maximization yields the same result regardless of these linear operations due to the associative property of the periodic likelihood function.

A sub-problem here is the initial classification when $\beta^i$ is not known, while $\alpha^i$ is known, we must find the most likely sequence of packets to belong to the process that has the inter-arrival time $\alpha^i$ for some $\beta^i$.
Then $g(T^i_{a}-T^i_{b},\alpha^i,\beta^i=0)$ is the likelihood that transmissions $a$ and $b$ belong to the same sequence with inter-arrival time $\alpha^i$. If we then let transmission $m$ be a newly received packet and $m^i$ be the last packet received by app $i$, then we can simply find $i$ by $\max_{i∈\{1:I_{\mathrm{apps}}\}}(g(T^i_m-T^i_{m^i},\alpha^i,\beta^i=0))$. Notice, that errors in the estimate of $\alpha$ will results in a error in the output of $g$, which increases as the distance between two timestamps increases, or in other words, labeling is more likely to be erroneous as ${p_\mathrm{o}}$ increases or after offline periods.

\section{Clustering} \label{sec:unlab}
When traffic parameters are unknown and the traffic is unlabeled by its parent application we need to perform clustering of the received traffic. In this section we propose a clustering algorithm that estimates clusters in a hierarchical manner based on the Lomb-Scargle periodogram and an online version, which assigns newly received transmissions to already known clusters in a greedy manner. %is needed in case the traffic is not labeled by its parent application and application number and parameters are unknown. 

% The sequence of received transmissions from device $x$ is the composite of the received transmissions from each application running on device $x$.
% %Since the composite stream is not easily sorted by an appID we indexing the data and parameters for each application by i we update our generating function from eq. \eqref{eq:peridic}: 
% \begin{align} \label{eq:peridiccomposite}
%     T^{x}_{\mathrm{RX}} = T^1_{\mathrm{RX}} \cup T^2_{\mathrm{RX}} \cup ... \cup T^{(I_{\mathrm{apps}}-1)}_{\mathrm{RX}} \cup T^{I_{\mathrm{apps}}}_{\mathrm{RX}}
% \end{align}

% \begin{align} \label{eq:peridiccomposite}
%     T_{m_i} = \beta_j + \alpha_j\cdot n_j + J_{m_j}
% \end{align}
% Here, we present a greedy hierarchical
Let the set of all unassigned traffic received by device $x$ be $T^{x*}_{\mathrm{RX}}$.The procedure for the hierarchical clustering method, Successive Periodicity Clustering (SPC) is described in Alg.~\ref{alg:SPC}. %, is:
% \begin{enumerate}
%     \item Hypothesize the parameters of an app, given $T^{x*}_{\mathrm{RX}}$.
%     \begin{itemize}
%         \item Estimate $\alpha^i$.
%         \item Check if the hypothesised $\alpha^i$ is significant based on \cite{Baluev2008}. If not, then SPC has finished.% Otherwise, continue.
%     \end{itemize}
%     \item Label data points $T^i_{\mathrm{RX}}$ and extract them from $T^{x*}_{\mathrm{RX}}$.
%     \begin{enumerate}
%         \item Label data points that best fit $\alpha_i$ and extract $\beta_i$.
%         %\item Save $app^i$ = {$T^i_{\mathrm{RX}}$, $\alpha_i$, $n_i$ and $o_i$}
%         \item Calculate KPIs based on $\alpha_i$, $\beta_i$ and $T^i_{\mathrm{RX}}$.
%         \item Remove $T^i_{\mathrm{RX}}$ from the set of unlabeled data. $T^{x*\text{\textbackslash} i}_{\mathrm{RX}} = T^{x*}_{\mathrm{RX}} \text{\textbackslash} T^i_{\mathrm{RX}}$
%     \end{enumerate}
%     \item Repeat from step 1 with $T^{x*}_{\mathrm{RX}}$ = $T^{x*\text{\textbackslash} i}_{\mathrm{RX}}$.
% \end{enumerate}
\begin{algorithm}[b]
\KwData{$T^{x}_{\mathrm{RX}}$}
\KwResult{\{$T^{i*}_{\mathrm{RX}}$, $\alpha^{i*}$ , $\beta^{i*}$ \}, $T^i_{\mathrm{RX}}\in T^{x}_{\mathrm{RX}}$}
\While{}{
Hypothesize $\alpha^{i*}$ for an app $i$, given $T^{x*}_{\mathrm{RX}}$.\\
\eIf{$\alpha^{i*}$ is significant}{ 
% Label data points $T^i_{\mathrm{RX}}$ and extract them from $T^{x*}_{\mathrm{RX}}$. \\
Label data points that best fit $\alpha^{i*}$ and extract $\beta^{i*}$ and $T^{i*}_{\mathrm{RX}}$.\\
Calculate KPIs based on $\alpha^{i*}$, $\beta^{i*}$ and $T^{i*}_{\mathrm{RX}}$.\\
Remove $T^{i*}_{\mathrm{RX}}$ from the set of unlabeled data. $T^{x*}_{\mathrm{RX}} = T^{x*}_{\mathrm{RX}} \text{\textbackslash} T^{i*}_{\mathrm{RX}}$
}{
return \{$T^{i*}_{\mathrm{RX}}$, $\alpha^{i*}$ , $\beta^{i*}$ \}, $T^i_{\mathrm{RX}}\in T^{x}_{\mathrm{RX}}$
}
% Update $T^{x*}_{\mathrm{RX}}$ = $T^{x*\text{\textbackslash} i}_{\mathrm{RX}}$.
}
\caption{Successive Periodicity Clustering (SPC)}\label{alg:SPC}
\end{algorithm}

% An example of an algorithmic implementation of the SPC procedure can be found in Algo.~\ref{alg:SPC} where 
NHM can be used for creating a hypothesis for $\alpha^i$. This is a robust approach as long as $\alpha^1 << \alpha^2 << ... << \alpha^{I_{\mathrm{apps}}}$ with an estimation error that increases greatly as $\dfrac{\alpha^i}{\alpha^{i+1}} \to 1$. However, it may often be the case that $\dfrac{\alpha^i}{\alpha^{i+1}} \approx 1$ so we will examine using the Lomb-Scargle periodogram for hypothesis creation instead.

Lomb in~\cite{Lomb1976} proposed a least-squares periodogram, which tested for the best fitting set of frequencies for a unevenly sampled time-series. In~\cite{Scargle1983}, Scargle proposed a generalised form of the classical Fourrier periodogram, which turns out to be equivalent to the least-squares fitting of sinusoids from~\cite{Lomb1976}. Hence, this periodogram was termed \emph{Lomb-Scargle periodogram} and~\cite{Baluev2008} treats its statistical properties, including a close-upper limit on the probability of falsely detecting a peak frequency in a data set comprised solely of noise. 
The Lomb-Scargle periodogram is well-known in some scientific communities, but was recently introduced to a wider audience in~\cite{VanderPlas_2018} by surveying works on and related to the Lomb-Scargle periodogram and lending conceptual intuitions. The Lomb-Scargle algorithm has been implemented in Python in the Astropy package~\cite{astropy:2013,astropy:2018}.

\subsection{Lomb-Scargle-based hypothesis creation for $\alpha$}
%The PSD-like characteristics found in the brute-force phase-folding methods inspires the usage of a Fourier method.
The classical Fourier periodogram requires evenly sampled data, but the Lomb-Scargle algorithm can find a PSD-like density for unevenly spaced times series~\cite{Lomb1976,Scargle1983,VanderPlas_2018}. 
%\rbs{Reference to SIC, which inspired the SPC algorithm?}

% \textbf{Lomb-Scargle, Fourier, Least Squares, }

Preprocessing of our data is required. %since Fourier methods are based on the analysis of sinusoids. 
We generate a series $V$ which is the same size as $T^{x*}_{\mathrm{RX}}$ and holds the value 1 for each of the timestamps. {Then we interpolate $V$ with 0's to permit sine-based analysis. We carry out this interpolation in a heuristic manner by finding places in $dT^{x*}_{\mathrm{RX}}$ larger than $dT^{x*}_{\mathrm{RX}}$ in an attempt to avoid interpolation between transmissions that are very close in time, i.e., to avoid giving credibility to very high frequencies.} %\rbs{R: double checking this in practise}

We must identify the frequency spectrum that is relevant for analysis. The size of the set of frequencies within the grid dictates both accuracy and computational effort, since 'false positive' local peaks in the periodogram are less likely to be found to be the global peak, but at the expense of evaluating the periodogram in more points.
The minimum detectable frequency will be a signal that completes one oscillation over the entire period of the data-set, $f_\text{min} = \dfrac{1}{T^{x*}_{M^{x*}}-T^{x*}_{1}}$. The spacing of the frequency grid, $\delta f = \dfrac{1}{n_0 \tau_\text{w}}$, depends on an oversampling factor, $n_0$. The higher $n_0$ is, the higher the chance that a peak frequency is not missed in the analysis. Typically, 8 is used~\cite{VanderPlas_2018}. 
The Nyquist limit does not always exist for the unevenly sampled Lomb-Scargle periodogram and in our case it will inevitably be very large since $f_{Ny} = 1/(2p)$, where $p$ is the largest value that all $T^{x*}_{\mathrm{RX}}$ can be written as an integer multiple of~\cite{lomb_fny}. Instead we shall use a heuristic to determine $f_{\max}$. %\PP{This is a strong sentence, good to have a reference}.
%  on our a priori  information about the generation functions of quasi-periodic apps. 
Say that $I_{\mathrm{apps}}$ applications are generating packets with rates $\alpha^i$ and there is no outage. In this case the mean observed inter-arrival time is always less than or equal to %\mathrm{mean}(dT^i_{\mathrm{RX}})
$\dfrac{1}{M^i-1}\sum_{{m^i}=1}^{M^i-1}(dT^i_{{m^i}}) \leq \min_{i\in\{1:I_{\mathrm{apps}}\}}(\alpha^i)$. %Changing the offsets, introducing outage or increasing the $\alpha$ of any application, but leaving one application at $\alpha_{min}$, we see that $\alpha_{mean} \geq \dfrac{\alpha_{min}}{L}$. 
 Then if we choose $f_{\max}$ such that 
 \begin{align}
     f_{\max} = \dfrac{L}{\dfrac{1}{M^{x*}-1}\sum_{{m^{x*}}=1}^{M^{x*}-1}(dT^{x*}_{{m^{x*}}})}
 \end{align}
 where L is scalar used to take into account the effect of outage. Given 50\% outage in a set of times one would expect to find a mean inter-arrival time that is twice as long. So assuming L = 2, our heuristic limit for the frequency grid should cover the maximum frequency of any observed app for ${p_\mathrm{o}} \leq 50\%$.
 
%  Note that in our previous example of L applications generating evenly spaced transmissions the transmissions will be identified as being generated by a single application with $IA^* = \dfrac{IA}{L}$. The outage estimated based on this will be the same so it doesnt matter
Let $Z = P(f)$ be the periodogram value at frequency $f$. In~\cite{Scargle1983} it was observed that the probability of observing a periodogram value less than Z in pure Gaussian noise can be expressed as \eqref{eq:FAP1}.
\begin{align} \label{eq:FAP1}
    P_{\mathrm{single}}(Z) = 1 - \exp{(-Z)}
\end{align}

A close-upper limit for the false alarm probability, $\mathrm{FAP}(Z)$, was found by Baluev in~\cite{Baluev2008}. We access this limit at the peak-value of the periodogram, $Z_0$, with significance, $\sigma$ and only accept the hypothesis $\alpha^* = 1/f_0$ if $\mathrm{FAP}(Z) < \sigma$.

\subsection{Labeling and collision resolution}
If a significant peak is detected we need to label data that fits the suggested $\alpha^{i*}$. Since up to $I_{\mathrm{apps}}$ apps have generated $T^x_{\mathrm{RX}}$ we must attempt to find the sequence of data points that is the best fit for our proposed application $i$. We do this by labeling transmissions, which fit coarsely and then sorting out incoherent false positive transmissions. Given a significant proposal for $\alpha^{i*}$ the procedure is:
\begin{enumerate}
    \item Mark transmissions that seem to be a good fit for $\alpha^{i*}$.
    \item Remove marks from transmissions that are incoherent with the rest of the labeled transmissions.
    \item Label marked transmissions as belonging to app $i$.
    \item Calculate KPIs based on app parameters and labeled packets.
\end{enumerate}

We calculate the fit between all times in all possible sequences by building a matrix $\textbf{IA} = \textbf{T}^{\textbf{x*}}_{\mathrm{\textbf{RX}}} \textbf{U} - \textbf{T}^{\textbf{x*}}_{\mathrm{\textbf{RX}}} \textbf{U}^T$ where $\textbf{U}$ is a vector of ones [1,1,1,...,1] of length $M$ and $\textbf{T}^{\textbf{x*}}_{\mathrm{\textbf{RX}}}$ is a vectorized version of the set of unlabeled transmissions. %Then we proceed to calculate the fit of every element in \textbf{IA} to the proposed $\alpha^{i*}$. 
We then proceed to estimate the fit, $\phi_m$, of each transmission to all other transmissions given $\alpha^{i*}$ as \eqref{eq:phi}.

\begin{align} \label{eq:phi}
\phi_m = \sum_{n=1}^M\big(g(\textbf{IA}_{(m,n)},\alpha^{i*}, \beta^i = 0)\big)    
\end{align}

%Then $\phi_m$ describes how well transmission $m$ fits with every other transmission. 
Now, a preliminary sorting of transmissions is performed by labeling only transmissions for which $\phi_m~>~\dfrac{1}{2 M^{x*}}\sum_{m=1}^{M^{x*}}(\phi_m)$. This coarse sorting of packets results in a fair amount of false positives in $T^{i*}_{\mathrm{RX}}$.
False positives can in many cases (especially when ${p_\mathrm{o}}$ is low) be identified as clearly incoherent transmissions that 'collide' in time with other transmissions in the set. The minimal period between packets is $\dfrac{\alpha^{i*}}{2}$ before a pair of transmissions are considered to collide. This is conditioned by $||J^i||<\dfrac{\alpha^{i*}}{4}$. Colliding pairs are resolved by checking the fit of the colliding packets and removing the worst fitting transmission from the set of transmissions for the new app. Given a collision at $T^x_{l}$ and $T^x_{l+1}$ we have that
% \begin{align} \label{eq:labelia}
%     S_{i,l} &= \sum_{m \in  T_{temp} \backslash m_{col}}(\cos(\dfrac{T_{m}-T_{m_{col}}}{\alpha^*}\cdot 2\pi))
% \end{align}
\begin{align} \label{eq:labels}
    C_{l\in\{1:M^i\}|app^i}:  \phi_{l} > \phi_{l+1}
\end{align}

Now we have estimated $T^i_{\mathrm{RX}}$ from the set of unlabeled data. Since the Lomb-Scargle method is only as accurate as the frequency grid we use in its analysis, we can attempt to use NHM on the estimated set $T^i_{\mathrm{RX}}$, to get better estimate for $\alpha^i$. Then we can compute KPIs for the application and finally remove the labeled traffic from $T^{x*}_{\mathrm{RX}}$ and attempt SPC again.% on the set of unlabelled traffic, $T^{x*\text{\textbackslash}i}$.

\subsection{Greedy online clustering}
SPC does batch processing and is relatively computationally heavy. An online version of the algorithm that minimizes the computational effort and can be used for real-time monitoring is warranted.
So we introduce a greedy online clustering (GOC) algorithm. Upon the reception of a new transmission $T^{x}_{M^X}$ GOC runs as described in Alg.~\ref{alg:GOC}.
% \begin{enumerate}
%     \item Estimate fit, $\phi^i$ for every previously known app $i$.
%     \item Check if $\max(\phi^i) > 0$:
%     \begin{itemize}
%         \item \textit{Yes} 
%         \begin{enumerate}
%             \item Add the transmission $T^{x}_{M^X}$ to app $T^{i}_{\mathrm{RX}}$.
%             \item Update KPIs
%         \end{enumerate}
%         \item \textit{No} 
%         \begin{enumerate}
%             \item Add the transmission to the set of unlabeled transmissions $T^{x*}_{\mathrm{RX}}$.
%             \item Run SPC on $T^{x*}_{\mathrm{RX}}$.
%         \end{enumerate}
%     \end{itemize}
% \end{enumerate}

\begin{algorithm}[t]
\KwData{$T^{x}_{M^X}$,\{$T^{i*}_{\mathrm{RX}}$, $\alpha^{i*}$ , $\beta^{i*}$ \}, $T^i_{\mathrm{RX}}\in T^{x}_{\mathrm{RX}}$}
\KwResult{\{$T^{i*}_{\mathrm{RX}}$, $\alpha^{i*}$ , $\beta^{i*}$ \}, $T^i_{\mathrm{RX}}\in T^{x}_{\mathrm{RX}}$}
Estimate the fit $\phi^{i*}$ of the new arrival $T^{x}_{M^X}$ for every previously known app $i$.\\
\eIf{$\max(\phi^i) > 0$}{
    Add the transmission $T^{x}_{M^X}$ to the app $j$ which had the best fit.\\
    Update the parameters, $\alpha^{i*}$ , $\beta^{i*}$.and the KPIs for $i$.
}{
    Add the transmission to the set of unlabeled transmissions $T^{x*}_{\mathrm{RX}}$. and run SPC on $T^{x*}_{\mathrm{RX}}$.
}
\caption{Greedy Online Clustering (GOC)}\label{alg:GOC}
\end{algorithm}

Once a new transmission is received, it should be checked if the reception time fits with the expected reception time of any known applications. 
GOC estimates $\phi^i~=~g(T^{x}_{M^X}~-~T^{i}_{M^i},\alpha^{i*},\beta^i~=~0)$ for all known $i$. Then assigns the transmission to $T^{i}_{\mathrm{RX}}$ if $\max_{i\in\{1:I_{\mathrm{apps}}\}}(\phi^i)$ is positive. Then the KPIs are updated for app $i$.

In case $\max_{i\in\{1:I_{\mathrm{apps}}\}}(\phi^i) \leq 0$ the new transmission is added to the set of unlabeled transmissions $T^{x*}_{\mathrm{RX}}$ and SPC is attempted.

\section{Probabilistic Performance} \label{sec:results}
In this section numerical results for the performance of the algorithms introduced throughout this paper are presented.  The performance is evaluated in terms of the accuracy of the outage estimation and offline state detection.

All the presented algorithms and the na\"{i}ve method presented in Sec. \ref{sec:naive} have been implemented in Python. %and can be found at \rbs{gitRef}. 
The implementations are based on functionality from the Numpy package~\cite{oliphant2006guide} and the Lomb-Scargle algorithm of the Astropy package~\cite{astropy:2013,astropy:2018}.
A parameter $\psi$ that reduces the computational requirements and enhances the accuracy of SPC and GOC has been introduced. $\psi$ ensures that SPC is only ever run when $T^{x*}_{\mathrm{RX}}$ is larger than $\psi$. Furthermore after labeling and collision resolution the size of the proposed set $T^{i}_{\mathrm{RX}}$ is checked and only accepted if it is larger than $\psi$. 

%\PP{Describe the data set used, emphasize that it is real data!}

The performance has been evaluated against network outage probabilities, ${p_\mathrm{o}} = [0, .1, .2, .3, .5]$. For each data point 1000 stochastic transmission sequences with jitter and outage have been generated. Outage has been randomly induced in the sequences by generating samples for $T^i_{\mathrm{RX}}$ and discarding $T^i_{\mathrm{m}^i}$ with probability ${p_\mathrm{o}}$ until the specified sample size was reached.
Sample sizes of 5, 10, 25, 50 and 100 transmissions have been used. %to illustrate behaviour both when only a few samples are available and converging behaviour for the estimates at a large number of samples.
The traffic parameters are drawn from the following distributions:
\begin{align}
\alpha^1 \sim U(100,200) \nonumber
\\\alpha^2 \sim U(1,5)\cdot\alpha^1 \nonumber
\\\beta^i \sim U(0,0.5)\cdot\alpha^i \nonumber
\\j^i_m \sim \exp(0.2)\cdot\dfrac{\alpha^i}{20} \nonumber
\end{align}

% All results are made using 1000 stochastic transmission sequences with Jitter that is sampled from a uniform distribution for each transmission under the condition that $J^i_m < \Big|\dfrac{\alpha^i}{100}\Big|$. 
% By considering the maximally possible jitter without creating aliasing the simulated results indicate the upper limit on the absolute estimation error.

\subsection{Network outage detection and estimation}
\begin{figure}[tb]
    \centering
    \includegraphics[width=1\columnwidth,trim={0 2cm 0 .5cm},clip]{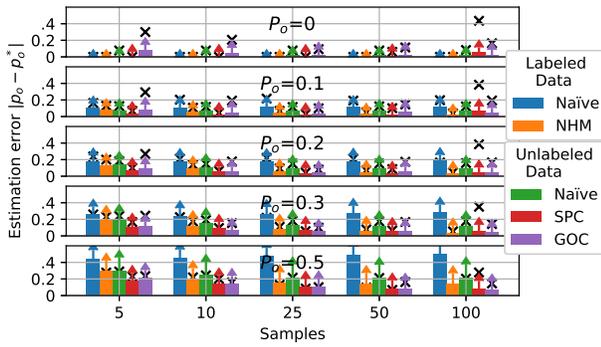}
    \caption{Absolute error in network outage estimation for NHM, GOC, SPC and the  na\"{i}ve approach. The  na\"{i}ve method is applied to both the labeled data-set and the unlabeled data-set ($\tau_w = 1500$). The bars indicate the mean absolute errors, the arrows indicate the size of the standard deviation and the 95th percentile is denoted by the black crosses. }
    \label{fig:clustering}
\end{figure}

The absolute estimation error, $|{p_\mathrm{o}}-{p_\mathrm{o}^*}|$, is plotted in Fig. \ref{fig:clustering} for NHM and the  na\"{i}ve method on a labelled sequence and SPC, GOC and the  na\"{i}ve method for unlabelled sequences. The labelled sequence is $T^1_{\mathrm{RX}}$, and the unlabelled sequence is $T^x_{\mathrm{RX}} = T^1_{\mathrm{RX}} \cup T^2_{\mathrm{RX}}$ and $\psi = \min(s_{\min},25)$ for SPC and GOC, where $s_{\min} = \min(M^1,M^2)$.

% \rbs{comments on results}
The na\"{i}ve approach works well when there is little to no uncertainty, but both the mean error and the variation in error increases rapidly as ${p_\mathrm{o}}$ increases. Notice that NHM has both a lower mean and less variance in the estimation error, especially as the number of available samples increase. Indeed, NHM has a $<$5\% estimation error at ${p_\mathrm{o}} = 0.3$ for 50 available samples. On unlabeled data-sets, The na\"{i}ve approach is found to perform a little better, which is expected due to the increased number of samples within the 'windowing' function of the na\"{i}ve approach. Still, both SPC and GOC outperforms the na\"{i}ve approach as ${p_\mathrm{o}}$ increases. We observe that SPC performs a little worse at 100 samples compared to 50 samples - this can be explained by the periodic labelling function performing worse over such a long period of samples, which is exaggerated when SPC then tries to find an 'imaginary' application to label transmissions that should have been assigned previously. In GOC, the inaccuracy of the periodic labelling function causes errors when only a few samples are available, here 'false positive' labels result in the secondary application not finding sufficient samples to satisfy $\psi$.

\subsection{Offline state detection}
\begin{figure}[tb]
    \centering
    \includegraphics[width=1\columnwidth,trim={0 1.2cm 0 0.5cm}]{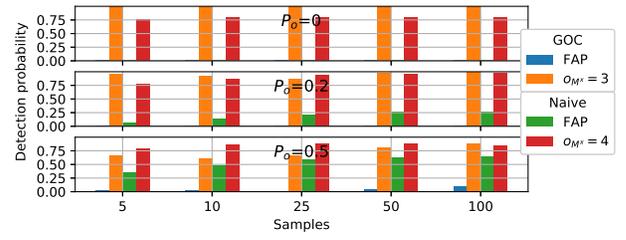}
    \caption{Probabilities of detecting offline-state in different conditions for the real-time clustering method, GOC, and the na\"{i}ve approach ($\epsilon = .25$, $\tau_w = 1500$). FAP denotes the false alarm probability. $o_{M^x}$ denotes the number of consecutively missed transmissions during the offline period.
    %and the composite sequence was generated such that $E[len(T^2)] = \dfrac{200}{3}$ and $\alpha^1 \leq \alpha^2 \leq 5 \cdot \alpha^1$.} % The offline app transmits between $2/3\cdot 100/5$  and $2/3\cdot 100$ packets.
    }
    \label{fig:offHist}
\end{figure}

In this scenario the device goes offline and we denote the number of transmissions since the device went offline, $o_M^x$. We define the false alarm probability (FAP) as the probability that an offline state is detected when the device is not offline. In Fig.~\ref{fig:offHist} the resulting probabilities are plotted. We observe that GOC identifies devices as offline correctly with a high probability. The probability of correctly identifying offline devices increases with the number of sampled reception times. The FAP for GOC increases with the network outage, as $\alpha^{i*}$ is more prone to be inaccurate, which is also the case when a very small number of samples are available before the device goes offline. Furthermore, when the network outage is high, the device is more likely to appear offline due to random outage in the received sequence $T^x_{RX}$, which affects the na\"{i}ve method significantly more than GOC.

% \rbs{True offline estimations as \% (of off). False alarm probabilities as \% (of on).}

% A histogram of 10k realisations of the outage estimation error of GOC can be found in Fig.~\ref{fig:offHist} for a a single device running two apps that goes offline. $T^x_{\mathrm{RX}}$ has been generated as for the clustering case, except here the device goes offline for a third of the total transmission period after a third of the total transmission period. 

% \begin{figure}
%     \centering
%     \includegraphics[width=.90\columnwidth]{OfflineInstance.png}
%     \caption{A single run showing the labeling of packet samples and outage estimation in real-time for GOC in a period, which includes an offline event for one of the applications.}
%     \label{fig:offInstance}
% \end{figure}

\subsection{Sampling and Computation time}

\begin{figure}[tb]
    \centering
    \includegraphics[width=1\columnwidth,trim={0 1.2cm 0 0.3cm},clip]{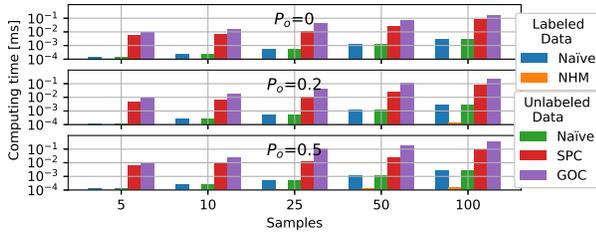}
    \caption{Mean computation time for NHM, SPC, GOC and the na\"{i}ve approach. Mind that the y-axis is logarithmic.}
    \label{fig:comptime}
\end{figure}

The total computation time of each algorithm is plotted in Fig.~\ref{fig:comptime}. The computation time of GOC is significantly larger than that of SPC, since GOC will attempt SPC multiple times as transmissions are received, however, one should note that the cmputing time of GOC when receiving transmissions from known apps is very low, near that of NHM. The computation time of NHM is significantly lower than that of any other method. The na\"{i}ve approach beats both clustering methods in terms of computation time. Still, in the special case that $\min(\alpha^i)$ is 1 hour for a group of devices, GOC supports initial analysis of up to 12240 devices in serial on a single thread of a 4.00~GHz i7-6700 if 100 samples are received as batch for each device, but processed in an online manner using GOC (worst-case). %Even more devices are supported after the devices' applications have been identified. 
In practice, the sampling time for MAR-P traffic \cite{3gppMAR} would be much greater, such that even more devices could be supported. %traffic would not arrive in simultaneously in batches of 100 samples, but instead as 1 sample at a time. It is evident that the sampling time for MAR-P traffic \cite{3gppMAR} would be much greater.
%The expected sampling time for S samples is given by $E[\tau_S ]=S\cdot\bigg(1+\dfrac{\mathrm{p_{o}}}{1-\mathrm{p_{o}}}\bigg)\cdot\alpha $.   
%It is evident that the computation time is much smaller than the sampling time, especially when $\alpha^i$ is large as can be expected for MAR-P traffic \cite{3gppMAR}.

% \begin{align} \label{eq:samptime}
% E[\tau_S ]=S\cdot\bigg(1+\dfrac{\mathrm{p_{o}}}{1-\mathrm{p_{o}}}\bigg)\cdot\alpha    
% \end{align}

% The expected sampling time for S samples is given by \*{eq:samptime}.
% The expected sampling time for S samples is given by $E[\tau_S ]=S\cdot\bigg(1+\dfrac{\mathrm{p_{o}}}{1-\mathrm{p_{o}}}\bigg)\cdot\alpha $.   
% It is evident that the computation time is much smaller than the sampling time, especially when $\alpha^i$ is large as expected in MAR-P traffic\cite{3gppMAR}.

% \begin{figure}
%     \centering
%     \includegraphics[width=.90\columnwidth]{SamplingTime.png}
%     \caption{Expected sampling time as a function of the outage.}
%     \label{fig:samptime}
% \end{figure}

% \subsection{Observation on jitter}
% \begin{figure}
%     \centering
%     \includegraphics[width=.90\columnwidth]{GOC.png}
%     \caption{Mean error of GOC as a function of the outage where $|J^i| < \alpha^i/100$ and the composite sequence was generated under the constraint that $\alpha_1 \leq \alpha_2 \leq 5 \cdot \alpha_1$.}
%     \label{fig:period_rcfe}
% \end{figure}

% \subsection{Application on real-world data} \label{sec:sve}
% Svebølle

% Monitoring and Scalability 

% (Kamstrup?)

% \subsection{Domain knowledge}
% \subsection{IoT devices}
% \begin{itemize}
% \item Multiple apps
% \item Weakly stationary
% \item Jitter
% \item Outage
% \end{itemize}

% \subsection{Analysis}

% \subsubsection{Heuristic Analysis}

% \subsubsection{Online Algorithms}

% \subsection{Results}

% \section{Lolland} \label{sec:lol}
% Monitoring and Scalability

% \subsection{Domain data}
% \subsection{IoT devices}
% \begin{itemize}
% \item On/Off
% \end{itemize}

% \subsection{Analysis}

% \subsubsection{Heuristic Analysis}

% \subsubsection{Online Algorithms}

% \subsection{Results}

% \section{Kamstrup}
% Provisioning

% \subsection{Domain data}
% \subsection{IoT devices}
% \begin{itemize}
% \item On/Off
% \end{itemize}

% \subsection{Analysis}

% \subsubsection{Heuristic Analysis}

% \subsubsection{Online Algorithms}

% \subsection{Results}

% \section{Application on real-world data} \label{sec:realdat}
\section{Case study: smart metering deployment} \label{sec:realdat}
This section presents monitoring results from using GOC on a data-set from a real-world wireless smart metering deployment. The data consists of timestamps, device IDs and IDs of the receiving concentrators for a total of 1048576 received transmissions from 4522 smart meters. The smart meters are deployed in a large area with 6 concentrators that provide coverage for all meters. Meters are connected to concentrators in a star-topology. %Each concentrator has two antennas, but we have only considered transmissions received at antenna \#0 for each concentrator.

% \begin{table}[b]
% \begin{tabular}{|l|l|l|l|l|l|}
% \hline
% Reception timestamp  & Meter ID & ColID & AntennaID %& RSSI
% & Packet size \\ \hline
% 2018-07-26  09:15:24 & M0       & C0    & 0         %& -93  
% & 33          \\ \hline
% 2018-07-26  09:15:45 & M1       & C1    & 0         %& -100 
% & 70          \\ \hline
% 2018-07-26  09:15:45 & M1       & C1    & 1         %& -88  
% & 70          \\ \hline
% \end{tabular}
% \vspace{0.01cm}
% \caption{Sample Data}
% \end{table}

Meters broadcast quasi-periodically to the concentrators at an inter-arrival time that is normally distributed around three hours.
It should be noted that meters, which do not receive an acknowledgement will include the data of any missed transmissions in their broadcast. Effectively this means that the app layer QoS is kept high even in poor link level conditions. Here, we monitor the network performance at the link level from the perspective of individual concentrators and the perspective of a centralised service gathering data from all concentrators. 

\begin{figure}[tb]
    \centering
    \includegraphics[width=.90\columnwidth,trim={0 0.4cm 0 0.2cm},clip]{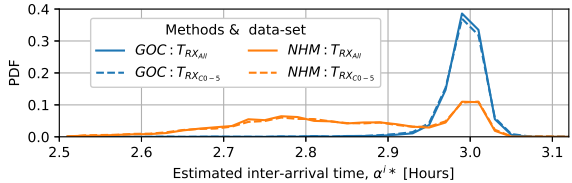}
    \caption{Distributions of estimated inter-arrival times from the perspectives of each concentrator and a centralised server.}
    \label{fig:kamia}
\end{figure}

The distribution of estimated $\alpha^{i*}$'s using GOC and NHM are plotted in Fig.~\ref{fig:kamia} for the perspective of a central server and as a joint distribution for all concentrators. Here we find that the estimated $\alpha^*i$'s are normally distributed around 3 hours for GOC, which is coherent with the deployment case. NHM on the other hand shows a fat tail in the distribution, which is not coherent with our prior knowledge of the transmission rate distribution. This inconsistency can be ascribed to the number of available data, noise in the data and the inaccuracy of NHM at increasingly higher network outages as seen in Fig. \ref{fig:clustering}.

\begin{figure}[tb]
    \centering
    \includegraphics[width=.90\columnwidth,trim={0 1.1cm 0 0.2cm},clip]{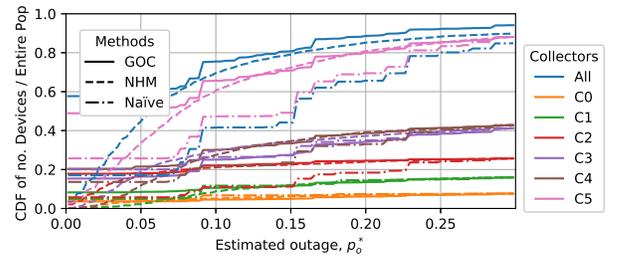}
    \caption{CDFs of estimated outages for each device as seen by each concentrator and a centralised server. The CDFs are evaluated against the entire population of devices in the deployment.}
    \label{fig:kamcolall}
\end{figure}

 %95.4\% of the devices are to at-least one concentrator at the link level.
%the  and 100\% of the devices are connected to at-least one concentrator at the link level. 
Overlapping concentrator coverage enhances the application level-performance; Using GOC, we find that 20.3\% of the meters are connected to at-least one concentrator without outage at the link-level, while the same measure is 54.8\% from the perspective of a centralised server.
The PDF of estimated outages for all meters from the perspective of the concentrators and the centralised server can be found in Fig.~\ref{fig:kamcolall}. Each given device is not necessarily in the range of each deployed concentrator, such that, as expected, we see varying connectivity levels for the different concentrators. The estimated outage at the link level as seen by the server is much lower than that of any individual concentrator. We have used a parameter of $\psi$ = 10 for the analysis. This means that 3.4\% of the devices represented in the data were not analysed due to being sampled less than 10 times while another 1.1\% did not exhibit clear periodicity due to having few samples and likely a relatively high outage. This group of devices are clear outliers in the deployment and warrant further investigation. %These 4.5\% of devices are the reason that Fig. \ref{fig:kamcolall} does not go to 1.

\begin{figure}[tb]
    \centering
    \includegraphics[width=.90\columnwidth,trim={0 .4cm 0 0.1cm},clip]{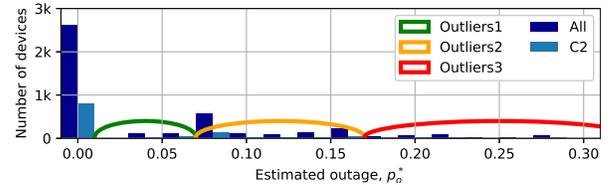}
    \caption{Histogram of estimated outages for devices in the deployment at the central server and concentrator C2. Devices can be grouped in spectra, which increase in width in proportion to the estimated outage, or equivalently, as reliability of the estimate goes down.}
    \label{fig:kamout}
\end{figure}

The distribution of the estimated outage by NHM and GOC converges after $p_0^* = 5\%$. Both methods identify roughly the same sets of devices for different groups of outliers. The na\"{i}ve method continually overshoots it's outage estimation. %, estimating that the number of expected transmissions within the data-set is far smaller than it actually is.
In Fig. \ref{fig:kamout} we take a closer look at the links experienced by the concentrator "C2" % (red, in the Fig.~\ref{fig:kamcolall})
and the links as experienced by the server. Here a histogram of the estimated outage at the link-level is depicted. In this way groups of outliers can be identified both at the application level and at the level of individual concentrators. 
In summary, the methods can quantify the coverage of concentrators in practice, which enables evaluation of their placement. The locations of concentrators and devices would be known by the smart metering company, making it relatively easy to assess spatial correlation of poorly performing devices. Installing a local concentrator is sensible in areas with many poorly performing devices, whereas antenna upgrades might be more cost beneficial for solitary, poorly performing devices. The methods can also help identifying Byzantine transmissions and seasonality in the outage. %The former by comparing the 'fitted' arrival-rate to the expected and the latter by using an appropriate windowing function on the real-time data trace.

%\PP{PP: We need to improve the selling argument here. Can you use 2-3 sentences to explain what this analysis offers to a smart metering company (without mentioning Kamstrup) that could not have been done without it? Which are the insights and which corrective actions could be taken? This could REALLY sell the paper. }

% \begin{figure}[tb]
%     \centering
%     \includegraphics[width=.90\columnwidth]{kam_outliers_all.png}
%     \caption{Identification of outliers for a single concentrator/antenna.}
%     \label{fig:kamout}
% \end{figure}

% \begin{figure}[tb]
%     \centering
%     \includegraphics[width=.90\columnwidth]{kam_outliers_k3(61029323).png}
%     \caption{Identification of outliers for a single concentrator/antenna.}
%     \label{fig:kamout}
% \end{figure}

\section{Conclusion} \label{sec:conc}
%In this paper we introduced a method for estimating the parameters of data generated by a single application, a method to label a sequence of transmissions by the application it belongs to and two clustering methods for doing both in composite transmission sequences from unknown applications. The latter were developed as a batch algorithm and online algorithm, respectively. Table \ref{tab:algapps} provides an overview of the algorithms and their applications. 

In this paper we introduced methods for passive detection in IoT networks in deployments with quasi-periodic reporting, such as smart-metering, environmental monitoring and agricultural monitoring. %The methods do not rely on topology information nor routing changes and are passive without any overhead. 
The methods are applicable in both mesh networks and LPWA and cellular networks, setting them apart from the state-of-art methods for fault detection in WSNs. %Table \ref{tab:algapps} provides an overview of the developed algorithms and their applications. 

% \begin{table}[]
% \resizebox{\columnwidth}{!}{
% \begin{tabular}{@{}ccccc@{}}
% \toprule
% Algorithm     & Classification & Regression & Clustering & Online \\ \midrule
% \begin{tabular}[c]{@{}c@{}}Sinusoidal\\likelihood\end{tabular} & x        &                                                                &            & x      \\
% NHM      &          & x                                                              &            & x      \\
% SPC      &          &                                                                & x          &        \\
% GOC      &          &                                                                & x          & x      \\ \bottomrule
% \end{tabular}
% }\caption{}\label{tab:algapps}
% \end{table}

The SPC and GOC algorithms were shown to perform well even at high outage for composite sequences, which makes them well suited for monitoring devices and networks in 'black-box' networks where the outage may be quite high. The cost of the utility and precision of these methods is computational effort, which was measured for a 4.0GHz i7-6700 CPU. Furthermore, the utility of NHM and GOC has been exhibited through a short analysis of real-world data from a smart meter deployment.

\section*{Acknowledgments}

This work was in part supported by a grant through FORCE Technology from the  Danish Agency for Institutions and Educational Grants, in part supported by the European Research Council (ERC) under the European Union Horizon 2020 research and innovation program (ERC Consolidator Grant Nr. 648382 WILLOW) and Danish Council for Independent Research (Grant Nr.8022-00284B SEMIOTIC).
%, (SEAS-NVE? if data used), (WILLOW?), any fond?

\midsepremove
% \begin{figure*}
\begin{table*}[tb]
{
\begin{tabularx}{\textwidth}{|c:X|c:X|c:X|}
\toprule 
Symbol                 & Meaning                                                                           & Symbol               & Meaning                                                                                    & Symbol                      & Meaning \\ \hline
$I^x_\mathrm{apps}$           & The number of applications running on device $x$.                                 & $m^i$                  & The index of the received packets from application $i$.
%$o^i_{m^i}$          & The cumulative number of missed receptions up to index $m$ from application $i$.        
&{$\tau_\text{w}$} & A time window. \\ 
$T^x_{\mathrm{TX}}$    & The set of all transmissions from device $x$.                                     & $T^i_m$              & The reception time of the $m$'th packet from application $i$.                                &$\eta^i_{m^i}$                &  The harmonic order of an observed inter-arrival rate. Used in NHM. \\
$T^{i}_{\mathrm{TX}}$  & The set of all transmissions from application $i$ on device $x$.                    & $\alpha^i$           & The period between transmission generation for quasi-peridic application $i$.                & $^*$                       &  The star denotes estimated values. \\
$T^x_{\mathrm{RX}}$    & The set of all received transmassions from device $x$.                            & $\beta^i$            & The offset in the period of application $i$.   & $f_{\min},f_{\max},\delta f$                     & The corner- and step-frequencies used in Lomb-Scargle. \\                                                 %& $L$ & A scalar used in SPC to compute the frequency grid. \\
$dT^{x}_{\mathrm{RX}}$ & The observed inter-arrival period for device $x$.      & $J^i_{m^i}$          & The E2E delay between generating a transmission to reception.                         & \textbf{IA} &        A matrix of all potential inter-arrival times for a set, $T^x_{\mathrm{RX}}$. \\
$T^{i}_{\mathrm{RX}}$  & The set of all received transmissions from application $i$.                       & $C_\mathrm{off}$            & A classification of being offline.%& $e(n^i)$             & A function that describes the period between transmissions for \rbs{event-driven application $i$.} 
& $\phi_m$                    & The likelihood of transmission $m$ being a part of a given application.\\
$dT^{i}_{\mathrm{RX}}$ & The observed inter-arrival period for application $i$. &${p_\mathrm{o}}$     &    The network outage   & $\Psi$                    & The minimal number of observations in an application.\\
$o^i_{m^i}$          & The cumulative number of missed receptions up to index $m$ from application $i$.
%$n^i$                  & The index of the transmissions from application $i$.  
& $k$                  &    The limit of consecutively missed transmissions before a device is classified as being offline.                                                                                        &$FAP(Z)$                    & A likelihood estimator for packet $m$ to belong to a specific application.\\%&  $Z$                & \\
$n^i$                  & The index of the transmissions from application $i$.  
%$m^i$                  & The index of the received packets from application $i$.
&      &                                                                                           &    &   \\ \bottomrule
%$M^i$                  & The total number of received packets from application $i$.                        & {$\tau_\text{w}$} & A time window.  &  & \\ \bottomrule % \sigma

\end{tabularx}
}
\caption{Index of notations and variables}
\label{tab:notes}
\end{table*}
% \end{figure*}
\midsepdefault

\ifCLASSOPTIONcaptionsoff
  \newpage
\fi



%\bibliographystyle{IEEEtran}
\bibliographystyle{IEEEtran}
\bibliography{bibref}

\begin{thebibliography}{10}
\providecommand{\url}[1]{#1}
\csname url@samestyle\endcsname
\providecommand{\newblock}{\relax}
\providecommand{\bibinfo}[2]{#2}
\providecommand{\BIBentrySTDinterwordspacing}{\spaceskip=0pt\relax}
\providecommand{\BIBentryALTinterwordstretchfactor}{4}
\providecommand{\BIBentryALTinterwordspacing}{\spaceskip=\fontdimen2\font plus
\BIBentryALTinterwordstretchfactor\fontdimen3\font minus
  \fontdimen4\font\relax}
\providecommand{\BIBforeignlanguage}[2]{{%
\expandafter\ifx\csname l@#1\endcsname\relax
\typeout{** WARNING: IEEEtran.bst: No hyphenation pattern has been}%
\typeout{** loaded for the language `#1'. Using the pattern for}%
\typeout{** the default language instead.}%
\else
\language=\csname l@#1\endcsname
\fi
#2}}
\providecommand{\BIBdecl}{\relax}
\BIBdecl

\bibitem{iotsurvey}
A.~{Al-Fuqaha}, M.~{Guizani}, M.~{Mohammadi}, M.~{Aledhari}, and M.~{Ayyash},
  ``Internet of things: A survey on enabling technologies, protocols, and
  applications,'' \emph{IEEE Communications Surveys Tutorials}, vol.~17, no.~4,
  pp. 2347--2376, Fourthquarter 2015.

\bibitem{NR5g}
S.~{Parkvall}, E.~{Dahlman}, A.~{Furuskar}, and M.~{Frenne}, ``Nr: The new 5g
  radio access technology,'' \emph{IEEE Communications Standards Magazine},
  vol.~1, no.~4, pp. 24--30, Dec 2017.

\bibitem{lpwans}
K.~Mekki, E.~Bajic, F.~Chaxel, and F.~Meyer, ``A comparative study of lpwan
  technologies for large-scale iot deployment,'' \emph{ICT Express}, vol.~5,
  no.~1, pp. 1--7, 2019.

\bibitem{nbiot}
S.~{Popli}, R.~K. {Jha}, and S.~{Jain}, ``A survey on energy efficient
  narrowband internet of things (nbiot): Architecture, application and
  challenges,'' \emph{IEEE Access}, vol.~7, pp. 16\,739--16\,776, 2019.

\bibitem{ltem}
R.~{Ratasuk}, N.~{Mangalvedhe}, D.~{Bhatoolaul}, and A.~{Ghosh}, ``Lte-m
  evolution towards 5g massive mtc,'' in \emph{2017 IEEE Globecom Workshops (GC
  Wkshps)}, Dec 2017, pp. 1--6.

\bibitem{netmonsurvey}
A.~{Mahapatro} and P.~M. {Khilar}, ``Fault diagnosis in wireless sensor
  networks: A survey,'' \emph{IEEE Communications Surveys Tutorials}, vol.~15,
  no.~4, pp. 2000--2026, Fourth 2013.

\bibitem{netmonsurvey2}
Z.~{Zhang}, A.~{Mehmood}, L.~{Shu}, Z.~{Huo}, Y.~{Zhang}, and M.~{Mukherjee},
  ``A survey on fault diagnosis in wireless sensor networks,'' \emph{IEEE
  Access}, vol.~6, pp. 11\,349--11\,364, 2018.

\bibitem{fuzzyframe}
\BIBentryALTinterwordspacing
M.~Nazari~Cheraghlou, A.~Khadem-Zadeh, and M.~a. Haghparast, ``A framework for
  optimal fault tolerance protocol selection using fuzzy logic oniot sensor
  layer,'' \emph{International Journal of Information \& Communication
  Technology Research}, vol.~10, no.~2, 2018. [Online]. Available:
  \url{http://ijict.itrc.ac.ir/article-1-326-en.html}
\BIBentrySTDinterwordspacing

\bibitem{pasDiag}
Y.~{Liu}, K.~{Liu}, and M.~{Li}, ``Passive diagnosis for wireless sensor
  networks,'' \emph{IEEE/ACM Transactions on Networking}, vol.~18, no.~4, pp.
  1132--1144, Aug 2010.

\bibitem{netmonbayes}
\BIBentryALTinterwordspacing
B.~C. Lau, E.~W. Ma, and T.~W. Chow, ``Probabilistic fault detector for
  wireless sensor network,'' \emph{Expert Systems with Applications}, vol.~41,
  no.~8, pp. 3703 -- 3711, 2014. [Online]. Available:
  \url{http://www.sciencedirect.com/science/article/pii/S0957417413009548}
\BIBentrySTDinterwordspacing

\bibitem{netmonar}
\BIBentryALTinterwordspacing
X.~Jin, T.~W.~S. Chow, Y.~Sun, J.~Shan, and B.~C.~P. Lau, ``Kuiper test and
  autoregressive model-based approach for wireless sensor network fault
  diagnosis,'' \emph{Wireless Networks}, vol.~21, no.~3, pp. 829--839, Apr
  2015. [Online]. Available: \url{https://doi.org/10.1007/s11276-014-0820-0}
\BIBentrySTDinterwordspacing

\bibitem{pdm}
R.~F. {Stellingwerf}, ``{Period determination using phase dispersion
  minimization.}'' \emph{Astrophysical Journal}, vol. 224, pp. 953--960, Sep
  1978.

\bibitem{Lomb1976}
N.~R. Lomb, ``Least-squares frequency analysis of unequally spaced data,''
  \emph{Astrophysics and Space Science}, vol.~39, no.~2, pp. 447--462, Feb
  1976.

\bibitem{Scargle1983}
J.~Scargle, ``Studies in astronomical time series analysis. ii - statistical
  aspects of spectral analysis of unevenly spaced data,'' \emph{The
  Astrophysical Journal}, vol. 263, 01 1983.

\bibitem{Baluev2008}
R.~V. {Baluev}, ``Assessing the statistical significance of periodogram
  peaks,'' \emph{Monthly Notices of the Royal Astronomical Society}, vol. 385,
  no.~3, pp. 1279--1285, April 2008.

\bibitem{VanderPlas_2018}
J.~T. VanderPlas, ``Understanding the lomb{\textendash}scargle periodogram,''
  \emph{The Astrophysical Journal Supplement Series}, vol. 236, no.~1, p.~16,
  may 2018.

\bibitem{6839076}
O.~{Al-Khatib}, W.~{Hardjawana}, and B.~{Vucetic}, ``Traffic modeling and
  optimization in public and private wireless access networks for smart
  grids,'' \emph{IEEE Transactions on Smart Grid}, vol.~5, no.~4, pp.
  1949--1960, July 2014.

\bibitem{3gppMAR}
3GPPP, ``Cellular system support for ultra-low complexity and low throughput
  internet of things (ciot),'' 3GPP, Technical report (TR) 45.820, 12 2015,
  version 13.1.0.

\bibitem{6740844}
A.~{Zanella}, N.~{Bui}, A.~{Castellani}, L.~{Vangelista}, and M.~{Zorzi},
  ``Internet of things for smart cities,'' \emph{IEEE Internet of Things
  Journal}, vol.~1, no.~1, pp. 22--32, Feb 2014.

\bibitem{6629817}
M.~{Laner}, P.~{Svoboda}, N.~{Nikaein}, and M.~{Rupp}, ``Traffic models for
  machine type communications,'' in \emph{ISWCS 2013; The Tenth International
  Symposium on Wireless Communication Systems}, Aug 2013, pp. 1--5.

\bibitem{astropy:2013}
{Astropy Collaboration} and T.~P. e.~a. {Robitaille}, ``{Astropy: A community
  Python package for astronomy},'' \emph{Astronomy and Astrophysics}, vol. 558,
  p. A33, Oct. 2013.

\bibitem{astropy:2018}
{Astropy Contributers} and A.~M. e.~a. {Price-Whelan}, ``{The Astropy Project:
  Building an Open-science Project and Status of the v2.0 Core Package},''
  \emph{Astronomical Journal}, vol. 156, p. 123, Sep. 2018.

\bibitem{lomb_fny}
L.~{Eyer} and P.~{Bartholdi}, ``{Variable stars: Which Nyquist frequency?}''
  \emph{Astronomy and Astrophysics, Supplement}, vol. 135, pp. 1--3, Feb. 1999.

\bibitem{oliphant2006guide}
T.~E. Oliphant, \emph{A guide to NumPy}.\hskip 1em plus 0.5em minus 0.4em\relax
  Trelgol Publishing USA, 2006, vol.~1.

\end{thebibliography}

\end{document}